\documentclass[]{jd04}    
\usepackage[]{graphicx}

\Title[SpectroWeb]{SpectroWeb: An Interactive Graphical Database of Digital Stellar Spectral Atlases} \Author{Lobel$^1$, A.} 
\Address{$^1$Royal Observatory of Belgium,\\
Ringlaan 3, Brussels, B-1180, Belgium \\
}{}

\ShortAuthors{Lobel}


\begin{document}
\maketitle

\begin{abstract}

\centerline{\em SpectroWeb is under development as an online interactive
repository}

\centerline{\em for identified spectral lines in spectral standard stars}.

SpectroWeb is an online maintained interactive graphical database of digital spectral atlases of 
spectral standard stars at 
http://spectra.freeshell.org . It is an efficient and user-friendly research tool for accurate analyses of stellar spectra observed with large spectral resolution, including the solar spectrum. The web-interface displays observed and theoretical stellar spectra, and comprehensively provides detailed atomic and molecular line information via user interaction. It fully integrates interactive spectrum visualization tools for the analysis, management, and maintenance of large volumes of spectral line-identification, -transition, and -property data. SpectroWeb 1.0 currently offers optical (3300-6800 \AA) flux normalized high-resolution spectra of Betelgeuse (M2), Arcturus (K1), The Sun (G2), Beta Aqr (G0), Procyon (F5), and Canopus (F0). The provided line identifications are based on state of the art spectrum synthesis calculations. The graphical database is under permanent development as an online repository of identified (absorption) lines in spectral standard reference stars, covering a broad range of stellar spectral types. Its object-oriented (Java) implementation offers future expansion capabilities to link and read stellar spectral atlases from various public internet sites. 

\end{abstract}

\section{Introduction}

SpectroWeb is a new online application that offers interactive high-resolution spectral atlases of astrophysical objects for the identification of spectral features. 
The demand for publicly available standard reference spectral atlases is steadily increasing with the fast improvements of spectral wavelength resolution and the quality by which stars of nearly all spectral types are being observed with modern spectrographs. Printed atlases of stellar spectra often only provide a small list of identified features without an assessment of the reliability of the spectral line identifications. Important information as to whether the line identifications are valid or not, or if they have been revised since publication, is often lacking. On the other hand, many public databases with spectral line information, that can be text queried online, are mostly based on theoretical calculations that have not been tested, or are difficult to test against observed stellar spectra. Users cannot readily assess if provided lists of spectral lines apply to their 
spectroscopic observations. The lists often contain 
line information that does not apply to an observed stellar spectrum because of unknown atmospheric formation conditions 
or chemical abundance differences with the solar values. Conversely, observed spectral features can often not be identified based on provided spectral line lists 
because the quality of the atomic and molecular 
line data is limited and requires further improvements 
and updates. The reader is invited 
to access the SpectroWeb interface online 
at {\tt spectra.freeshell.org} using sufficiently 
large computer screen resolutions (a minimum of 1024 by 768 pixels is recommended) for the detailed presentation of graphical data combined with textual data. Figures 1-3 show screencopies of the SpectroWeb interface using color coding for the combined graphics and text displays. 
The web interface is accessible with any type of modern Internet browser in 
which the Java interpreter has been activated.

\section{SpectroWeb Development Status}

SpectroWeb is an efficient online tool to directly assess the quality of spectral line identifications by comparing high-quality observed stellar spectra with state-of-the-art computed spectra. The web interface enables users to select 10 or 25 \AA\, wide spectral regions-of-interest from an interactive list of observed wavelengths listed to the left of the spectrum displayed
in Fig. 1. The continuum normalized observed spectra ({\it black drawn line}) and computed spectra ({\it red drawn line}) are overplotted and marked with numbered spectral line identifications ({\it at arrows}) in case the central line flux differs by more than two percent from the flux level of the stellar continuum. Users can zoom in on the displayed spectral regions by selecting smaller regions of interest using the left mouse button. Atomic, molecular, and Earth spectral line identifications in the region of interest can be selected 
from the top menu and are listed in an interactive table to the right of the spectrum. Each identified spectral line can then be selected from the tables to list corresponding detailed atomic and molecular line transition and line property information
below the visualized spectrum. 

\begin{figure}[!ht]
\centerline{\includegraphics[width=30pc]{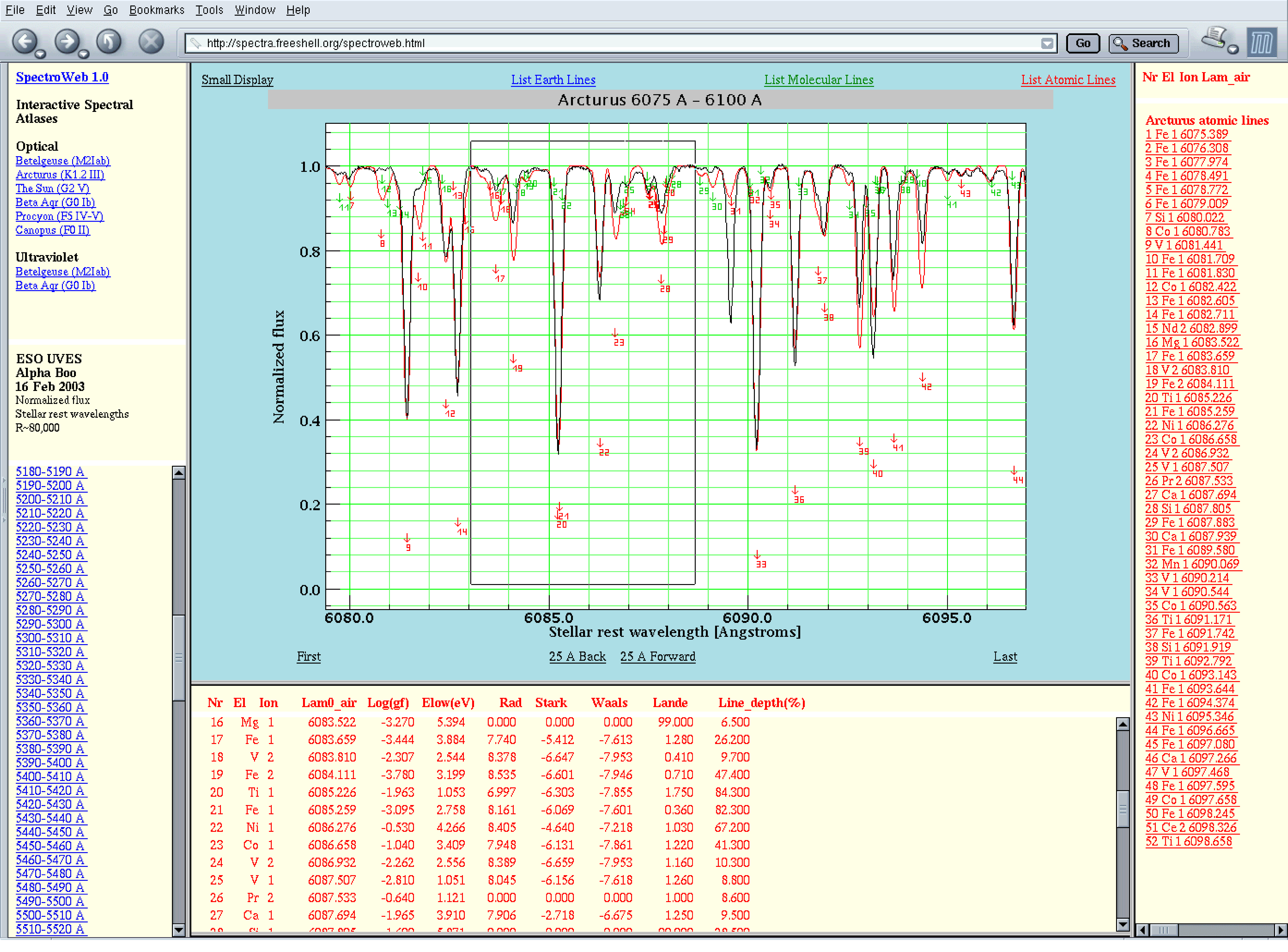}}
\caption{The SpectroWeb interface is shown for the continuum normalized spectrum of Arcturus over $\sim$18 \AA\, 
around 6088 \AA. The numbered arrows in the spectrum 
correspond to identified atomic ({\it red labels}) and molecular ({\it green labels}) absorption lines. The atomic lines are listed to the right of the spectrum. The table color is related to the color of the numbered line markers and alters when \underline{List Atomic Lines} ({\it red color}), \underline{List Molecular Lines} ({\it green color}), and \underline{List Earth lines} ({\it blue color}) identifications are selected from the top menu. Users can switch between various target stars in the blue left-hand table and select the same region-of-interest from a list of provided wavelengths for detailed comparisons of line strengths, widths, or line overlap. Fast paging through the spectral regions is possible with the \underline{Back} and \underline{Forward} buttons in the menu below the spectrum. A \underline{Large Display} or \underline{Small Display} can be selected depending on the user's screen resolution. For the atomic lines SpectroWeb currently lists the oscillator strength ({\bf log gf}), the energy of the lower transition level ({\bf Elow}), the four line damping constants ({\bf Rad.}, {\bf Stark}, {\bf Waals}, {\bf Land\'{e}}), and the computed normalized line depth without instrumental broadening ({\bf Line depth}). Weak CN lines of the AX band system can also be identified (i.e. labeled Nrs. {\bf 41} \& {\bf 42} around 6095 \AA), which can be listed by selecting \underline{List Molecular Lines} in the top menu.  }
\label{fig1}
\end{figure}

\begin{figure}[!ht]
\centerline{\includegraphics[width=30pc]{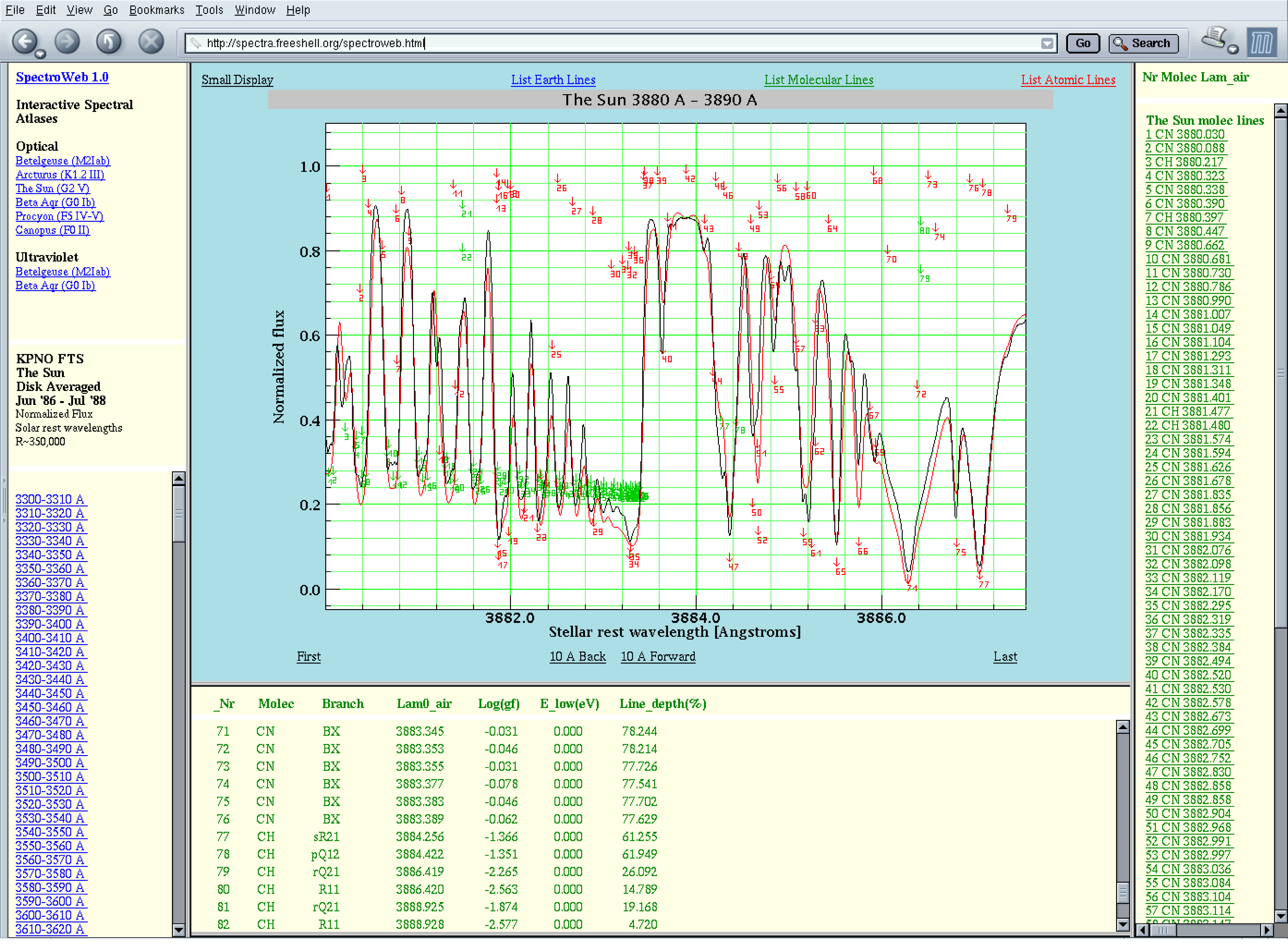}}
\caption{  SpectroWeb is shown for the solar spectrum around the strong molecular bandhead of CN at 3883 \AA. Individual CN lines are marked with green arrows. The bandhead starts at line Nr. 
{\bf 76} and degrades toward shorter wavelengths. The green table below the spectrum lists the lines from the BX band system of CN with computed central line depths. The molecular lines in the spectrum longward of the CN bandhead are due to CH. The atomic line Nr. {\bf 71}, with broad damping wings and marked with a red arrow, is due to Fe~{\sc i} 3886.28 \AA, which can be listed by selecting \underline{List Atomic Lines} in the top menu.  }
\label{fig2}
\end{figure}

The SpectroWeb 1.0 database currently includes observed and theoretical high-resolution spectra of five bright cool stars and The Sun, available in international ground-based telescope archives; 
Betelgeuse ($\alpha$ Ori, $T_{\rm eff}$=3500 K), 
Arcturus ($\alpha$ Boo, 4500 K), Beta Aqr (5500 K), 
Procyon ($\alpha$ CMi, 6500 K), and Canopus ($\alpha$ Car, 
7500 K). The increase of about 1000 K between these stars towards earlier spectral types causes large changes in the optical high-resolution spectra due to large changes of the stellar atmospheric ionization balance. For the coolest stars line blending strongly increases towards shorter wavelengths, resulting in a large decrease of the local continuum flux level to below the stellar continuum level. For M supergiant Betelgeuse, however, the optical spectrum is dominated by molecular opacity and mainly due to titanium-oxide. 

\begin{figure}[!ht]
\centerline{\includegraphics[width=30pc]{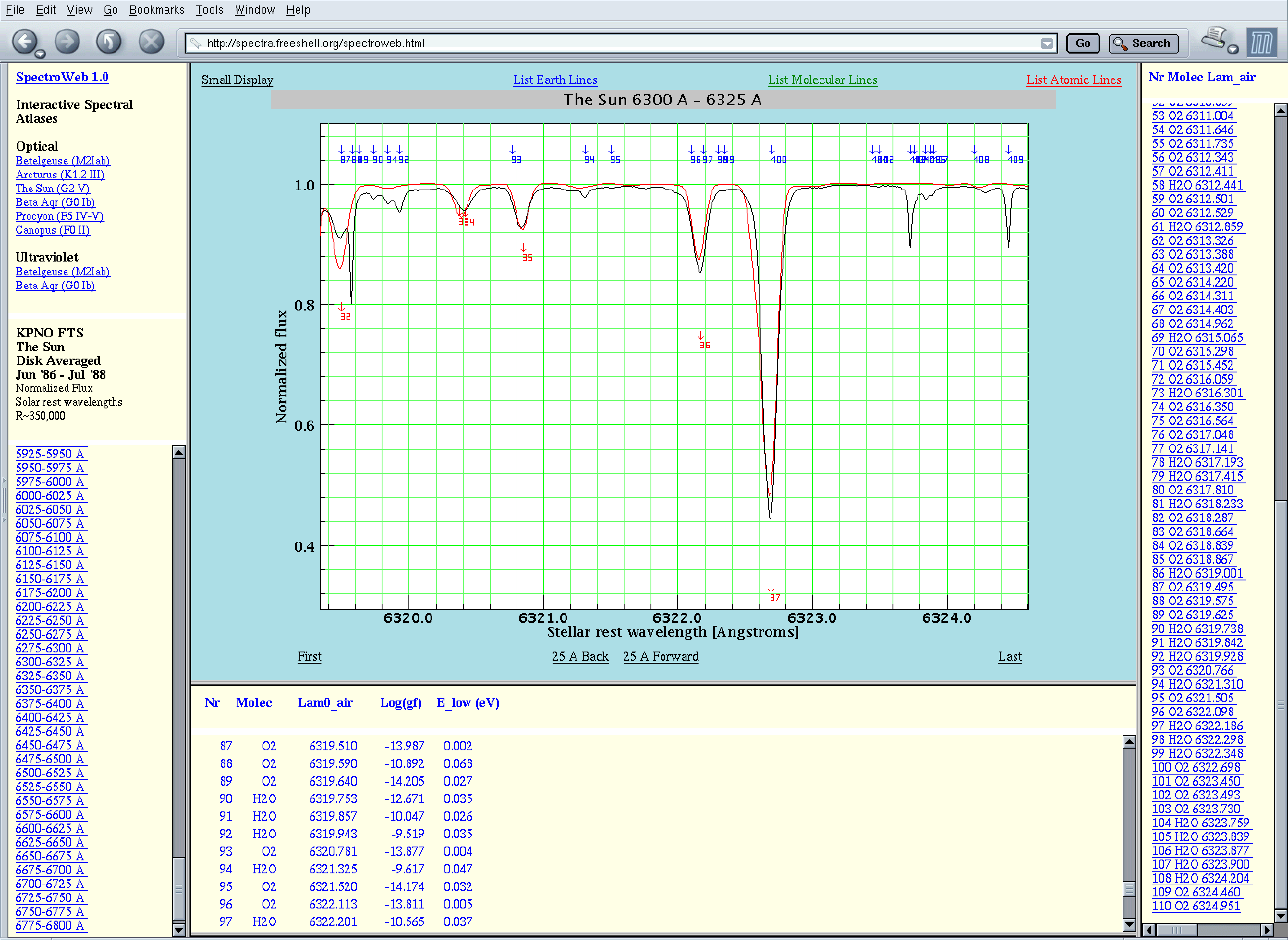}}
\caption{ SpectroWeb is zoomed in on the solar spectrum around the strong atomic line of Fe~{\sc i} 6322.69 \AA\, (marked with red arrow Nr. {\bf 37}). The solar continuum normalized flux level in this wavelength region is very close to unity and weak Earth lines can be observed in the spectrum. For example, the lines marked Nrs. {\bf 90}, {\bf 91}, {\bf 92}, and {\bf 94}, with blue arrows in the upper part of the spectrum panel are due to water vapor, and listed in the blue table below it. The lines marked with blue arrows Nrs. {\bf 103} \& {\bf 109} are deeper 
($\sim$10\%) and due to telluric $\rm O_{2}$. The data for these lines can be listed by scrolling down the blue table below the spectrum, or by selecting the individual lines \underline{\bf 103 O2 6323.730} or \underline{\bf 109 O2 6324.460} in the blue right-hand table.  }
\label{fig3}
\end{figure}

The high-resolution optical atlases in SpectroWeb 1.0 currently cover the wavelength range between 3300 \AA\, and 6800 \AA. The observed spectra of these bright cool stars are of good quality and the theoretical spectra can identify nearly all of the medium-strong and strong Fraunhofer lines which often blend together with weaker lines. Important broad diagnostic lines such as H$\alpha$, H$\beta$, and Na~$D$ 
are well-matched, and most of the spectral features that blend with the line wings are identified. The theoretical spectra are computed using a grid of stellar atmosphere models (Kurucz 1994; Castelli \& Kurucz 2003). The lists of atomic lines for metals to compute the spectra are available online (Kurucz \& Bell 1995; Smith et al. 1996). Other atomic lines are from the Vienna Atomic Line Database (VALD-2; Kupka et al. 1999). Diatomic molecules are also included. The theoretical spectra have been computed for solar chemical abundance values so far. They currently exclude telluric lines due to water vapor and $\rm O_{2}$ in Earth's atmosphere. The strongest $\rm H_{2}O$ and $\rm O_{2}$ lines are marked with blue arrows. The theoretical spectra are convolved with a Gaussian filter to simulate the instrumental profile. Many spectral lines therefore blend together, which can sometimes be determined from asymmetries in observed spectral features. The current linelists to calculate the spectra are however incomplete and several observed spectral features (mostly weak absorption lines) require further improvements of the atomic input data, or have yet to be identified. More stars, wavelength regions, and spectral line information will be added to SpectroWeb. The goals of its object-oriented implementation are to link and read stellar spectral atlases in one database that is served from various public websites worldwide using a standard fast interactive display.

\section{Summary}
SpectroWeb is an online repository of identified lines
observed in high-resolution spectra of spectral standard stars
including The Sun. The web interface currently offers optical 
and near-UV spectra and will be expanded to also incorporate 
near-infrared wavelength regions. The database is permanently 
updated with improved line transition information for identified spectral features in observed high-quality stellar spectra utilizing advanced theoretical spectral synthesis calculations. 

\noindent
{\it Acknowledgements:}

This work has been supported by the Belgian 
Federal Science Policy - Terug-keermandaten.

\end{document}